# Diffusion of a magnetic skyrmion in 2-dimensional space

Yoshishige Suzuki[1,2,†], Soma Miki[1,2,†], and Eiiti Tamura[1,2]
1 Graduate School of Engineering Science, Osaka University, Toyonaka, Osaka 560-8531, Japan
2 Center for Spintronics Research Network (CSRN), Graduate School of Engineering Science, Osaka University, Osaka 560-8531, Japan

**Abstract** Two-dimensional magnetic skyrmions are particle-like magnetic domains in magnetic thin films. The kinetic property of the magnetic skyrmions at finite temperature is well described by the Thiele equation, including a stochastic field and a finite mass. In this paper, the validity of the constant-mass approximation is examined by comparing the Fourier spectrum of Brownian motions described by the Thiele equation and the Landau-Lifshitz-Gilbert equation. Then, the 4-dimensional Fokker-Planck equation is derived from the Thiele equation with a mass-term. Consequently, an expression of the diffusion flow and diffusion constant in a tensor form is derived, extending Chandrasekhar's method for Thiele dynamics.

*: Corresponding author:

    Yoshishige SUZUKI: suzuki-y@mp.es.osaka-u.ac.jp

    Soma MIKI: miki@spin.mp.es.osaka-u.ac.jp

    Osaka University,

    Graduate School of Engineering Science,

    Department of Materials Engineering Science,

    Toyonaka, Osaka 560-8531, Japan

†: Those authors are equally contributed to the paper.

**Introduction**

Magnetic skyrmions are particle-like magnetic domains characterized by a topological number [1]. In the magnetic thin films without space inversion symmetry, interfacial Dzyaloshinskii-Moriya interaction stabilizes 2-dimensional magnetic skyrmions [2]. Because of its small size (several nanometers to several micrometers) and topological protection, a skyrmion is a candidate to realize 3-dimensionally integrated race-track memory [3]. Recently, utilization of the Brownian motion of the skyrmion in stochastic computing [4] and in Brownian computing [5] was proposed. Those modern applications require a deeper understanding of the Brownian motion and diffusion of semi-classical skyrmions [6, 7, 8, 9].

The finite temperature dynamics of the skyrmions are well described by the Landu-Lifschitz-Gilbert (LLG) equation, including a stochastic field and spin-torque terms [10]. In addition, its particle-like nature allows us to use the Thiele equation [11] to describe the dynamics of the skyrmion. On the application of the latter equation, it has been suggested that one needs to include frequency-dependent mass and gyro-damping [6]. In addition, to understand its dynamics, the chiral property should be adequately treated.

This paper provides an analytical expression of the diffusion flow of a skyrmion. The flow is described in the Fokker-Planck equation derived from the Thiele equation, including a stochastic field and constant mass. For this purpose, the accuracy of the constant-mass approximation is examined by comparing the Fourier spectra of the Brownian motion obtained from the Thiele equation to those from the LLG micromagnetic simulation. Eventually, a usual diffusion-flow expression with diffusion constants in a tensor form [8] is obtained for the time scale longer than the relaxation time of the skyrmion.

**Results**

The Thiele equation is derived [11] from the LLG equation, assuming that the magnetic texture moves without changing its shape. In this paper, the generalized Thiele equation [10], which includes a mass-term and a stochastic field, is utilized to include the effects of deformation of the skyrmion and finite temperature,

$$m\frac{d\mathbf{v}(t)}{dt} = -\alpha D \mathbf{v}(t) - G \mathbf{e}_z \times \mathbf{v}(t) + \mathbf{f}(t) \qquad (1)$$

This paper considers a single skyrmion in ferromagnetic films extended in the *x-y* plane where the skyrmion-skyrmion interaction is neglected. $\mathbf{v}(t) \equiv d\mathbf{x}(t)/dt$ is the velocity of the skyrmion at time *t*. $\mathbf{x}(t)$ is the position of the skyrmion. The first term in the right-hand side (RHS) of the equation is a damping term including Gilbert damping factor α and dissipation dyadic *D*. The 2nd term in RHS is the gyro-term including z-component of the gyro-vector *G*. Here, for simplicity, we use an expression in eq. (1) that is correct for the uniform film with *C*$_{3v}$ or higher symmetry. The gyro-damping term [6] does not exist in the original Thiele equation and is small in the ref. [6], is omitted to avoid difficulty to treat it with the Fokker-Planck equation caused by its amplification ability. $D$ and $G$ are expressed as follows in terms of the magnetization distribution $\mathbf{M}(\mathbf{x})$.

$$\begin{cases} D = \left(\dfrac{\mu_0 h M_s}{-\gamma}\right)\dfrac{1}{M_s^2}\int_{-\infty}^{+\infty} d^2 x \dfrac{\partial \mathbf{M}}{\partial x}\cdot\dfrac{\partial \mathbf{M}}{\partial x} \\ G = 4\pi\left(\dfrac{\mu_0 h M_s}{-\gamma}\right) n_{Sk} \end{cases} \qquad (2)$$

$M_s$ is saturation magnetization, $h$ film thickness, $\mu_0 = 4\pi\times 10^{-7}\ H/m$ vacuum magnetic permeability, and $\gamma < 0$ the gyromagnetic ratio of the host ferromagnetic material. $n_{Sk}$ is the skyrmion number (winding number) defined by the following formula.

$$n_{Sk} = \dfrac{1}{4\pi}\dfrac{1}{M_s^3}\int_{-\infty}^{+\infty} d^2 x\, \mathbf{M}\cdot\left(\dfrac{\partial \mathbf{M}}{\partial x}\times\dfrac{\partial \mathbf{M}}{\partial y}\right) \qquad (3)$$

The skyrmion number is a topological quantity and is an integer. In this paper, $\alpha D$ and $G$ are treated as positive and real constant parameters, respectively.

At the left-hand side (LHS) of Equation (1), the acceleration term with constant mass $m$ is added to the Thiele equation to describe the inertia of the skyrmion [6]. At the RHS, a stochastic force term, $\mathbf{f}(t)$, is added so as to describe the dynamics under the thermal noise. $\mathbf{f}(t)$ has a time correlation of $\langle f_i(t)f_j(t+\tau)\rangle = 2k_B T\alpha D\delta_{ij}\delta(\tau)$ [12]. Here, $k_B$ is the Boltzmann constant, and $T$ is the absolute temperature.

To verify the constant-mass approximation, the thermodynamical property of the skyrmion described by the generalized Thiele equation (1) is compared with that obtained from the LLG micromagnetic simulation. First, a skyrmion is formed at zero temperature, whose structure is shown in Figure 1. The parameters employed for the simulation are $\alpha$=0.03, $M_s$=580 kA/m, $h$=1.2 nm, $\gamma/(2\pi)$=-28 GHz/T, uniaxial anisotropy constant Ku=0.9 MJ/m³, exchange stiffness constant $A_{ex}$=27 pJ/m, anti-symmetric exchange constant $J_{DMI}$=-4.5 mJ/m, space mesh size 1 nm × 1 nm, and time step 10 fsec. From the shape of the skyrmion, $D$ and $G$ are evaluated to be 7.3×10⁻¹⁴ kg/s and -5.0×10⁻¹⁴ kg/s, respectively. The radius and width of the transition region of the skyrmion are 10.7 nm and 5.3 nm, respectively.

In figure 2, a trajectory of the skyrmion at $T$=300 K obtained by the LLG simulation is shown. Brownian motion of the skyrmion is observed. In figure 3, the power spectrum of the velocity of the Brownian motion is displayed by the blue dots. Flat frequency dependence is obtained below 45 GHz. The peak at around 45 GHz reflects the dynamics caused by the finite mass.

The formal solution of the generalized Thiele equation (1) is,

$$\begin{cases} \mathbf{v}(t) = \int_{-\infty}^{+\infty} d\omega\, \hat{g}(\omega)\mathbf{f}(\omega)e^{-i\omega t} \\ \mathbf{f}(\omega) = \dfrac{1}{2\pi}\int_{-\infty}^{+\infty} dt\, \mathbf{f}(t)e^{i\omega t} \end{cases}, \qquad (4)$$

where $\hat{g}(\omega)$ is the green function of the equation,

$$\begin{cases} \hat{g}(\omega) \equiv \dfrac{1}{m}(\hat{\gamma} - i\omega\hat{I})^{-1} \\ \hat{\gamma} \equiv \dfrac{1}{m}\begin{pmatrix} \alpha D & -G \\ G & \alpha D \end{pmatrix} \equiv \begin{pmatrix} \gamma & -\Gamma \\ \Gamma & \gamma \end{pmatrix} \end{cases}. \qquad (5)$$

Here $\hat{I}$ is the 2x2 identical matrix. $\hat{g}(\omega)$ has its poles at $\omega_\pm = -i\alpha D/m \pm G/m$. Since $\mathbf{f}(t)$ has a white spectrum, the power spectrum of the velocity of the Brownian motion is expressed as,

$$\langle v_x(t)^2 \rangle \equiv \dfrac{2k_B T \alpha D}{2\pi} \sum_{k=1,2} \int_{-\infty}^{+\infty} d\omega\, g_{xk}(\omega)g_{xk}(-\omega). \qquad (6)$$

The fitting curve using expression (6) is shown in figure 2 by a red curve. For the fitting, $\alpha D$ and $G$ are fixed to the values obtained by the LLG simulation. While $m$ is employed as a fitting parameter. The fitting results $m=1.8\times10^{-25}$ kg. The obtained mass is at least one order smaller than that obtained experimentally ($8\times10^{-22}$ kg for a skyrmion with a radius of around 87 nm) [13]. Almost perfect fitting up to 45 GHz suggests the validity of the constant-mass approximation for the dynamics slower than the gyro-frequency, $G/m$.

The above verified constant-mass approximation allows us to construct a Fokker-Planck equation, which describes the time evolution of the probability density of the stochastic valuables, i.e., $\mathbf{x}$ and $\mathbf{v}$. From $d\mathbf{x}(t)/dt = \mathbf{v}(t)$ and Equation (1), flows of the probability density, $p(\mathbf{x},\mathbf{v},t)$, in the position-space, $\mathbf{J}_x(\mathbf{x},\mathbf{v},t)$, and the velocity-space, $\mathbf{J}_v(\mathbf{x},\mathbf{v},t)$, are,

$$\begin{cases} \mathbf{J}_x(\mathbf{x},\mathbf{v},t) = \mathbf{v}(t)p(\mathbf{x},\mathbf{v},t) \\ \mathbf{J}_v(\mathbf{x},\mathbf{v},t) = -\left(\hat{\gamma}\mathbf{v} + \gamma\dfrac{k_B T}{m}\nabla_v\right)p(\mathbf{x},\mathbf{v},t) \end{cases}, \qquad (7)$$

where $p(\mathbf{x},\mathbf{v},t)$ is the probability distribution function, $\nabla_x \equiv (\partial/\partial x, \partial/\partial y)$ and $\nabla_v \equiv (\partial/\partial v_x, \partial/\partial v_y)$. The first line in Equation (7) is the direct result of $d\mathbf{x}(t)/dt = \mathbf{v}(t)$. The first term in the second line in Equation (7) is obtained from $d\mathbf{v}(t)/dt = -\hat{\gamma}\mathbf{v}(t)$. And the second term is the diffusion term. Probability conservation results in following Fokker-Planck equation in 4-dimensional phase space,

$$\frac{\partial p(\mathbf{x},\mathbf{v},t)}{\partial t} = -(\nabla_x \cdot \mathbf{J}_x + \nabla_v \cdot \mathbf{J}_v) = -\left(\nabla_x \cdot \mathbf{v} - \nabla_v \cdot \left(\hat{\gamma}\mathbf{v} + \gamma \frac{k_B T}{m}\nabla_v\right)\right) p(\mathbf{x},\mathbf{v},t), \quad (8)$$

Equations (7) and (8) are in agreement with the general Langevin equation [14]. Applying Chandrasekhar's method [15] to the Thiele equation, we obtain the solution of the Fokker-Planck equation,

$$\begin{cases} P_0(\tilde{\mathbf{x}},t) = \dfrac{1}{\sqrt{(2\pi)^4 \det Q(t)}} \mathrm{Exp}\left[-\dfrac{1}{2}\tilde{\mathbf{x}}^t Q^{-1}(t)\tilde{\mathbf{x}}\right] \\ Q^{-1} = \dfrac{m}{k_B T(h_1(t) - \det[\hat{\gamma}^{-1}])}\begin{pmatrix} \hat{I} & -(\hat{\gamma}^{-1})^t \\ -\hat{\gamma}^{-1} & h_1(t)\hat{I} \end{pmatrix}, \\ h_1(t) = \det[\hat{\gamma}^{-1}]\left(2\gamma t - 2\dfrac{\gamma^2 - \Gamma^2}{\gamma^2 + \Gamma^2}\right) \end{cases} \quad (9)$$

where $\tilde{\mathbf{x}}^t = (\mathbf{x}^t, \mathbf{v}^t)$ is the 4-columns position-velocity vector. Eq. (9) is the solution for the initial conditions with $\mathbf{x}(0)=0$ and Boltzmann distribution of the velocity. In addition, transient terms with exponential decay are neglected for simplicity. Equation (9) is correct for $t \gg \gamma^{-1}$. The derivation of the Q-matrix, including the transient term, is shown in the Appendix.

The convolution expresses the solution with the initial position density $n_0(\mathbf{x})$,

$$\begin{cases} P(\mathbf{x},\mathbf{v},t) = \int_{-\infty}^{+\infty} d^2 x_0 n_0(\mathbf{x}_0) P_0(\mathbf{x} - \mathbf{x}_0, \mathbf{v}, t) \\ n(\mathbf{x},t) = \int_{-\infty}^{+\infty} d^2 v\, P(\mathbf{x},\mathbf{v},t) \\ \mathbf{J}(\mathbf{x},t) = \int_{-\infty}^{+\infty} d^2 v\, P(\mathbf{x},\mathbf{v},t)\mathbf{v} \end{cases} \quad (10)$$

Here expressions for the position density $n(\mathbf{x},t)$ and the skyrmion flow density $\mathbf{J}(\mathbf{x},t)$ at time $t$ are also shown. Gaussian type probability allows us to use the following relation, which is valid for $t \gg \gamma^{-1}$.

$$-\frac{k_B T}{m}\int_{-\infty}^{+\infty} d^2 v\, \nabla P_0(\mathbf{x},\mathbf{v},t) = \hat{\gamma}\int_{-\infty}^{+\infty} d^2 v\, \mathbf{v}\, P_0(\mathbf{x},\mathbf{v},t). \quad (11)$$

Alternating eq. (11) in (10), we obtain,

$$\mathbf{J}(\mathbf{x},t) = -\hat{\mathfrak{D}} \nabla_x n(\mathbf{x},t), \qquad (12)$$

where

$$\hat{\mathfrak{D}} \equiv \frac{k_B T}{m} \hat{\gamma}^{-1} = \frac{k_B T}{(\alpha D)^2 + G^2} \begin{pmatrix} \alpha D & G \\ -G & \alpha D \end{pmatrix}, \qquad (13)$$

is the diffusion constant. The eq. (12) is the usual expression of the flow of the diffusion except that the diffusion constant is now a tensor form. The expression (13) is equivalent to that in ref. [8] except for the sign of the off-diagonal elements because of a different definition of the footing.

As is described in Equation (8), the initial transient dynamics occur with a time constant $\tau_{relax} = m/(\alpha D)$. Therefore, Equations (9, 11,12) and following Equations (14-16) are valid for the time scale longer than $\tau_{relax}$. In our simulation, $\tau_{relax}$ is around 80 ps. The anomaly at the beginning of the Brownian motion was once discussed by Ornstein and Fürth [16-18]. We also discuss it for the skyrmions using the exact solution given in the Appendix.

The continuity equation for the flow expressed in Equation (12) is,

$$\frac{\partial}{\partial t} n(\mathbf{x},t) = \mathfrak{D}_{xx} \Delta n(\mathbf{x},t), \qquad (14)$$

and is called the diffusion equation.

Equation (12) clearly shows that the existence of deflected diffusion flow with respect to the gradient of the density, which we call "gyro-diffusion". Since the gyro-diffusion has no divergence, the resulted equation of diffusion (Eq. 14) does not include off-diagonal elements of the diffusion constant. Therefore, the gyro-diffusion is hidden in the observation of the time evolution of the concentration.

The gyro-diffusion is, however, observable if we look at the system microscopically. By observing the Brownian motion, one may examine the velocity and position of particles and evaluate a velocity-position correlation function [8]. It is straightforward to show that

$$\langle \mathbf{v}(t) \otimes \mathbf{x}(t) \rangle_{ij} \equiv \langle v_i(t) x_j(t) \rangle = \int_{-\infty}^{+\infty} d^2x d^2v \, v_i x_j \, p(\mathbf{x},\mathbf{v},t)$$
$$= -\frac{k_B T}{m} \sum_{k=1}^{2} \gamma^{-1}_{ik} \int_{-\infty}^{+\infty} d^2x d^2v \, x_j \frac{\partial}{\partial x_k} n(\mathbf{x},t) = \frac{k_B T}{m} \gamma^{-1}_{ij} = \mathfrak{D}_{ij} \qquad (15)$$

Therefore, an experimental observation in the velocity-position correlation function may provide a measure in the gyro-diffusion constant.

To check the existence/nonexistence of the perpetual rotational flow in the equilibrium[19], a central force in a harmonic potential, i.e., $-k\mathbf{x}$ is added to the Thiele equation (1). The new term results in a drift flow in Equation (12).

$$\mathbf{J}(\mathbf{x},t) \cong \hat{\mathcal{D}} \frac{-\nabla_x \left(\frac{1}{2}k\mathbf{x}^2\right)}{k_B T} n(\mathbf{x},t) - \hat{\mathcal{D}} \nabla_x n(\mathbf{x},t). \tag{16}$$

Under the thermal equilibrium, $n(\mathbf{x},t)$ is equal to the Boltzmann distribution. And a cancellation between drift flow and diffusion flow results in zero local-net-flow even though the chiral symmetry of the system is broken. As it is known, the radial drift flow brings skyrmions to the bottom of the potential, but diffusion flow is opposite, and they cancel each other. For the rotational flow, the same happens. Diffusion flow shows a natural sense of the rotation direction, but the drift flow is the opposite. Under any strength of the harmonic potential, the cancellation is perfect, and there is no local net flow in the thermal equilibrium. It is apparent that the situation is the same for skyrmions bounded by the arbitrary potential. Therefore, there is no perpetual flow in the skyrmion system under the thermal equilibrium. The above discussion is an analog to the theory of diffusion on the classical 2-dimensional electron system under a magnetic field [20], which obeys the Bohr-Van Leeuwen theorem [21].

**Conclusions**

In this paper, a simple formula that connects the diffusion flow of the 2-dimensional magnetic skyrmion to the Thiele equation was derived by means of the Fokker-Planck equation. The obtained formula shows the existence of gyro-diffusion flow. The method to observe the gyro-diffusion was also discussed. The result is applicable not only to the magnetic skyrmions but also to the chiral particles such as classical electrons in the magnetic field.

**Appendix**

The generalized Thiele equation generates the following samples of the position and velocity as functions of the elapsed time under given initial conditions.

$$\begin{cases} \mathbf{x}(t) = \mathbf{x}_d(t) + \mathbf{x}_s(t) \\ \mathbf{v}(t) = \mathbf{v}_d(t) + \mathbf{v}_s(t) \end{cases}. \tag{A1}$$

Here, subscripts *d* and *s* denote "deterministic" and "stochastic" parts of the trajectories. Stochastic parts are obtained by the following integrals of the stochastic force.

$$\begin{cases} \mathbf{v}_s(t) = \int_0^t \xi_v(t-t_1)\mathbf{f}(t_1)dt_1 \\ \mathbf{x}_s(t) = \int_0^t \xi_x(t-t_1)\mathbf{f}(t_1)dt_1 \\ \xi_x(t_1) = \int_{t_1}^t \xi_v(t-t_2)dt_2 \end{cases} \quad (A2)$$

Since $\mathbf{x}_s(0) = \mathbf{v}_s(0) = 0$, initial conditions are satisfied by the deterministic parts. Extending Chandersekhar's method [15] to the generalized Thiele equation (1), The $Q_0$-matrix and the probability density for the given initial condition are determined by the following equation,

$$\begin{cases} \tilde{\boldsymbol{\mu}}^t Q_0 \tilde{\boldsymbol{\mu}} = \int_0^t dt_1 \left(\boldsymbol{\mu}_x{}^t \xi_x(t_1) + \boldsymbol{\mu}_v{}^t \xi_v(t_1)\right)\left(\xi_x(t_1)^t \boldsymbol{\mu}_x + \xi_v(t_1)^t \boldsymbol{\mu}_v\right)\dfrac{2k_B T}{m}\gamma \\ P(\mathbf{x},\mathbf{v},t|\mathbf{x}(0),\mathbf{v}(0)) = (2\pi)^{-4}\int_{-\infty}^{+\infty} d^4\mu \, \mathrm{Exp}\left[-i\boldsymbol{\mu}\cdot(\mathbf{x}-\mathbf{x}_d(t)) - i\boldsymbol{\mu}_v\cdot(\mathbf{v}-\mathbf{v}_d(t)) - \dfrac{1}{2}\tilde{\boldsymbol{\mu}}^t Q_0 \tilde{\boldsymbol{\mu}}\right] \end{cases}, \quad (A3)$$

where $\tilde{\boldsymbol{\mu}}^t = (\boldsymbol{\mu}_x{}^t, \boldsymbol{\mu}_v{}^t)$ is the 4-column parameter vector for the Fourier integral. Here, $Q_0$ is independent of the initial condition. Q-matrix in the main text is obtained by taking the thermal average for $\mathbf{v}(0)$ and $\mathbf{x}(0) = 0$ in (A3), i.e.,

$$P_0(\mathbf{x},\mathbf{v},t) = \int_{-\infty}^{+\infty} d^2 v(0) P_{\mathrm{Th}}(\mathbf{v}(0)) P(\mathbf{x},\mathbf{v},t|\mathbf{x}(0)=0,\mathbf{v}(0)) = \left((2\pi)^4 \sqrt{\det Q}\right)^{-1} \mathrm{Exp}\left[-\dfrac{1}{2}\tilde{\boldsymbol{\mu}}^t Q^{-1} \tilde{\boldsymbol{\mu}}\right], \quad (A4)$$

where $P_{\mathrm{Th}}(\mathbf{v})$ is the Boltzmann distribution.

An explicit expression for Q-matrix, which is valid from $t=0$ is,

$$Q^{-1} = \dfrac{1}{\Xi\Psi - \Theta^2 - \Phi^2}\begin{pmatrix} \Psi\hat{I} & -(\Theta\hat{I} - \Phi\hat{E}) \\ -(\Theta\hat{I} + \Phi\hat{E}) & \Xi\hat{I} \end{pmatrix}$$

$$\begin{cases} \Xi = \dfrac{k_B T}{m}\dfrac{2}{\gamma^2+\Gamma^2}\left(\gamma t - \dfrac{\gamma^2-\Gamma^2}{\gamma^2+\Gamma^2} + \left(\dfrac{\gamma^2-\Gamma^2}{\gamma^2+\Gamma^2}\cos\Gamma t - \dfrac{2\gamma\Gamma}{\gamma^2+\Gamma^2}\sin\Gamma t\right)e^{-\gamma t}\right) \\ \Psi = \dfrac{k_B T}{m} \\ \Theta = \dfrac{k_B T}{m}\dfrac{\gamma + (-\gamma\cos\Gamma t + \Gamma\sin\Gamma t)e^{-\gamma t}}{\gamma^2+\Gamma^2} \\ \Phi = \dfrac{k_B T}{m}\dfrac{\Gamma - (\Gamma\cos\Gamma t + \gamma\sin\Gamma t)e^{-\gamma t}}{\gamma^2+\Gamma^2} \end{cases}, \quad (A5)$$

where $\hat{E} = \begin{pmatrix} 0 & 1 \\ -1 & 0 \end{pmatrix}$.

**Acknowledgments** The authors thank Dr. K. Saito of Keio University and Dr. Y. Imai of Nagoya University for their valuable comments on statistical physics. This work was supported by JSPS KAKENHI (S) Grant Numbers JP20H05666 and JST CREST Grant Number JP JPMJCR20C1, Japan.

**Competing interests statement** There are no competing interests relating to this research.

**Yoshishige Suzuki**: Formal analysis, Investigation, Supervision **Soma Miki**: Formal analysis, Investigation **Eiiti Tamura**: Conceptualization

Figure captions

Fig. 1 Shape of the skrmion obtained from the micromagnetic simulation with Landau-Lifshitz-Gilbert (LLG) equation. Z-axis shows the direction cosine of the magnetization along with film normal.

Fig. 2 Trajectory of the skyrmion Brownian motion during 50 ns. Simulations were done using Landau-Lifshitz-Gilbert (LLG) equation with the time step of 10 fs. The plot was done using data points taken every 1 ps.

Fig. 3 Power spectrum density of the velocity of the Brownian motion shown in Fig. 3 (blue dots). Red dots show the theoretical power spectrum obtained using the Thiele equation with parameters extracted from the LLG equation.

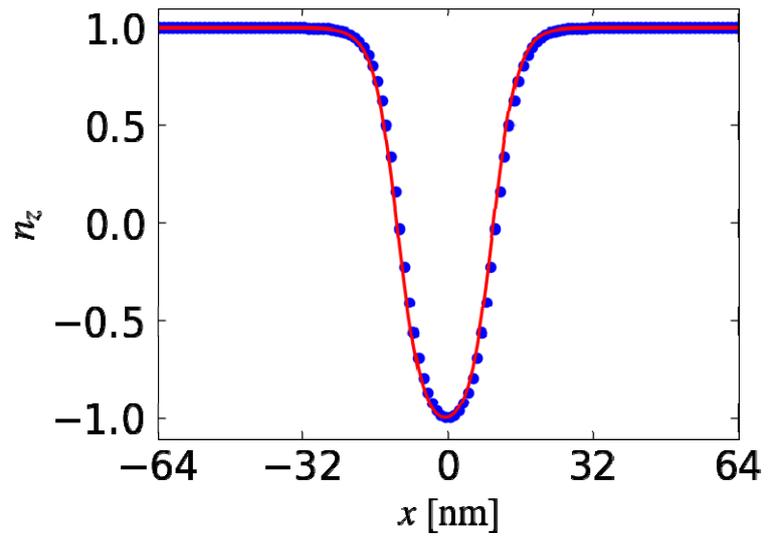

Fig. 1 (Color)

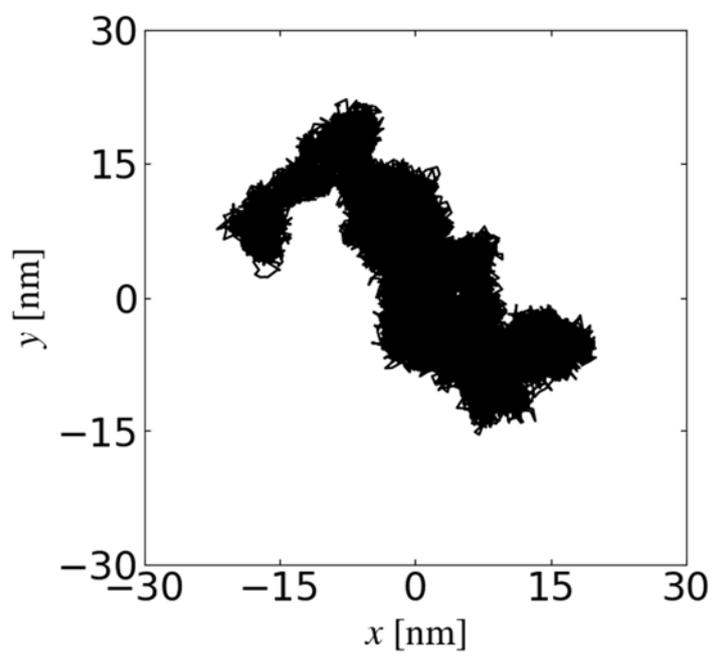

Fig. 2

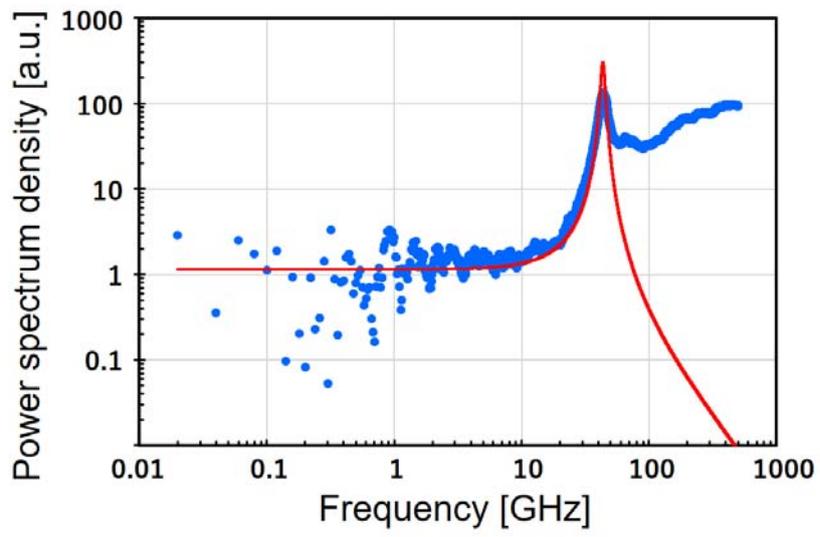

Fig. 3 (Color)